\documentclass[aps,prb,twocolumn]{revtex4}
\usepackage{epsfig,graphicx}

\begin{document}

\title{\bf Evolutionary approach for finding the atomic structure of
\\  steps on stable crystal surfaces}

\author{Ryan M. Briggs$^{\dagger, \ddagger}$ and Cristian V.
Ciobanu$^{\dagger ,}$ \footnote{Corresponding author; Phone
303-384-2119; Fax 303-273-3602; Email: cciobanu@mines.edu}  }

\affiliation{$\dagger$Division of Engineering, Colorado School of
Mines, Golden, Colorado 80401
\\
$\ddagger$ Division of Engineering and Applied Science, California
Institute of Technology, Pasadena, California 91125 }

\begin{abstract}
The problem addressed here can be concisely formulated as follows:
given a stable surface orientation with a known reconstruction and
given a direction in the plane of this surface, find the atomic
structure of the steps oriented along that direction. We report a
robust and generally applicable variable-number genetic algorithm
for step structure determination and exemplify it by determining
the structure of monatomic steps on Si(114)-$2\times 1$. We show
how the location of the step edge with respect to the terrace
reconstructions, the step width (number of atoms), and the
positions of the atoms in the step region can all be
simultaneously determined.
\end{abstract}
\maketitle

One-dimensional (1-D) nanostructures presently show tremendous
technological promise due to their novel and potentially useful
properties. For example, gold chains on stepped silicon surfaces
\cite{553-Au-Spain,553-Au-Wisc} can have tunable conduction
properties, rare earth nanowires and bismuth nanolines have
unusual straightness and length \cite{bowler-review} and can thus
be useful as nanoscale contacts on chips or as templates for the
design of other novel structures. The structure of step edges on
silicon surfaces is of key interest, for it can help trigger a
step-flow growth mode \cite{bcf} useful for preparing high-quality
wafers. Understanding the formation, properties, and potential
applications of these intriguing 1-D nanostructures requires
knowledge of the atomic positions of various possible adsorbate
species, as well as of the location of the silicon atoms at the
step edges.

The determination of the atomic configuration at crystal surfaces
is a long-standing problem in surface science. With the invention
and widespread use of the scanning tunelling microscope (STM) our
understanding of crystal surfaces (in particular of surface
reconstructions) has progressed immensely. Nevertheless, since the
STM acquires information about the local density of states and not
about the atomic positions, one still needs to find structural
models that correspond to the images acquired. The problem becomes
more challenging in the case of one-dimensional nanostructures on
semiconductor surfaces (steps, surface-supported nanowires, atomic
or molecular chains), because the step and nanowire structures may
not be readily inferred from STM images given the possibility of
having either flat supporting surfaces or vicinal ones with
single- or multiple-layer steps. Even for a given direction and
step height, the number of possible structures is daunting and
their identification is tedious because it currently relies on
relaxing ad-hoc structures that may or (more often) may not end up
corresponding to the experiments. As seen in the case of Au
\cite{553-Au-Spain,553-Au-Wisc} or Ga \cite{Ga-wires-si112} on Si
surfaces, one needs to propose a large number of atomic models and
then check if they have sufficiently low formation energies in
order to ultimately identify or predict them as the actual
physical nanostructures.

Motivated by the need to find good candidates for one-dimensional
structures on surfaces, we have set out to develop a global search
procedure that creates and selects atomic models based on their
formation energy. To fix ideas, we address here the problem of
predicting the atomic structure of steps along a given direction
on a stable surface with a known reconstruction. While we focus on
the case of straight steps on high-index semiconductor surfaces,
we point out that the evolutionary procedure with variable atom
numbers described below is generally applicable for finding the
structure of any surface-supported 1-D nanostructure, provided
that suitable (i.e. fast and reasonably transferable) interatomic
potentials are available.

\begin{figure}
 \includegraphics[width=8.1cm]{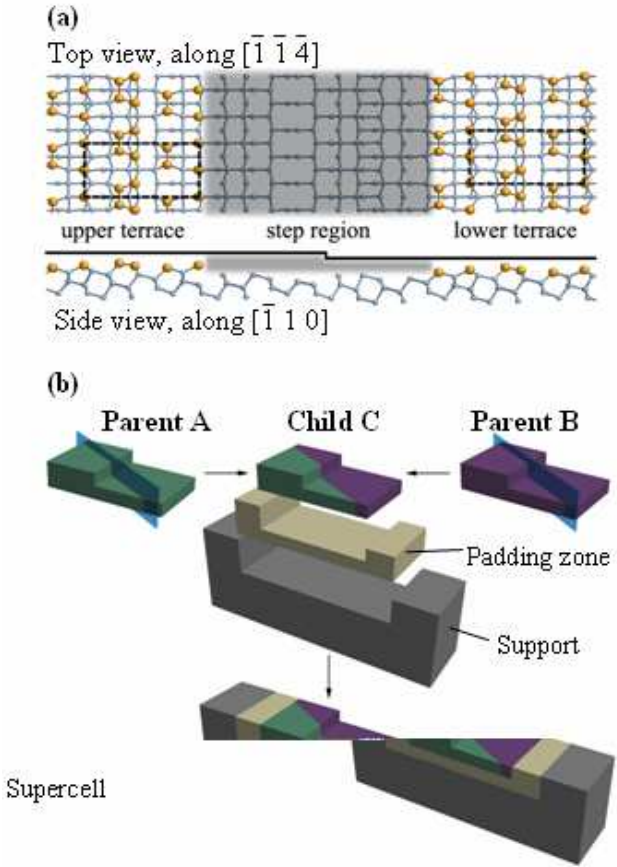}
\caption{(Color online)(a) Step region (shaded) for which the
number of atoms, their positions, and the location of the step
edge are to be determined. The step region, which corresponds to
the $[{\overline 1}10]$-down configuration, is surrounded by
reconstructed terraces with the surface unit cell marked by the
dashed rectangles. (b) Crossover operation used in the genetic
algorithm to search the configuration space. The step energy
calculations involve relaxing a padding zone (in addition to the
step region) with the supporting zone fixed. }
\label{boundary-mating}
\end{figure}

The particular 1-D systems chosen for this study are steps on
stable high-index semiconductor surfaces [e.g., Si(114), Si(5512),
Si(113)],\cite{Si114-prl, 5512-science,113} which have a
fundamental importance in addition to the practical one alluded
above. Due to the lack of a robust approach for proposing and
sorting step models, the pioneering study \cite{chadi} of steps on
Si(001) was followed by only a few reports of step structures on
other semiconductor surfaces.
\cite{gaas-steps,shengbai-gaas,105-bunching-SS} Focusing on the
case of steps on the Si(114)-$2\times1$ surface, experiments show
that straight steps form along the $[{\overline 1}10]$ and the
$[22{\overline 1}]$ directions,\cite{Si114-apl} which are
precisely the directions of spatial periodicity of the
reconstructed Si(114) unit cell.\cite{Si114-prl} For each of the
two directions we can define two types of steps (up and down), and
for each step type there are two relative positions of the
reconstructed unit cells on the upper and lower terraces:  one in
which the unit cells on terraces are in registry (normal) and
another in which they are offset (shifted). A crystallographic
analysis of the Si(114) surface shows that out of the four
possible terrace configurations for each step direction there can
be only two that are topologically distinct. The configurations
that we have to address are therefore four, denoted here by
$[{\overline 1}10]$-down, $[{\overline 1 }10]$-up, $[22{\overline
1}]$-normal, and $[22{\overline 1}]$-shifted.
{Fig.~\ref{boundary-mating}a} illustrates the $[{\overline
1}10]$-down configuration, while the remaining ones are described
in note \onlinecite{geosimcell}.

\begin{figure}
 \includegraphics[width=8.3cm]{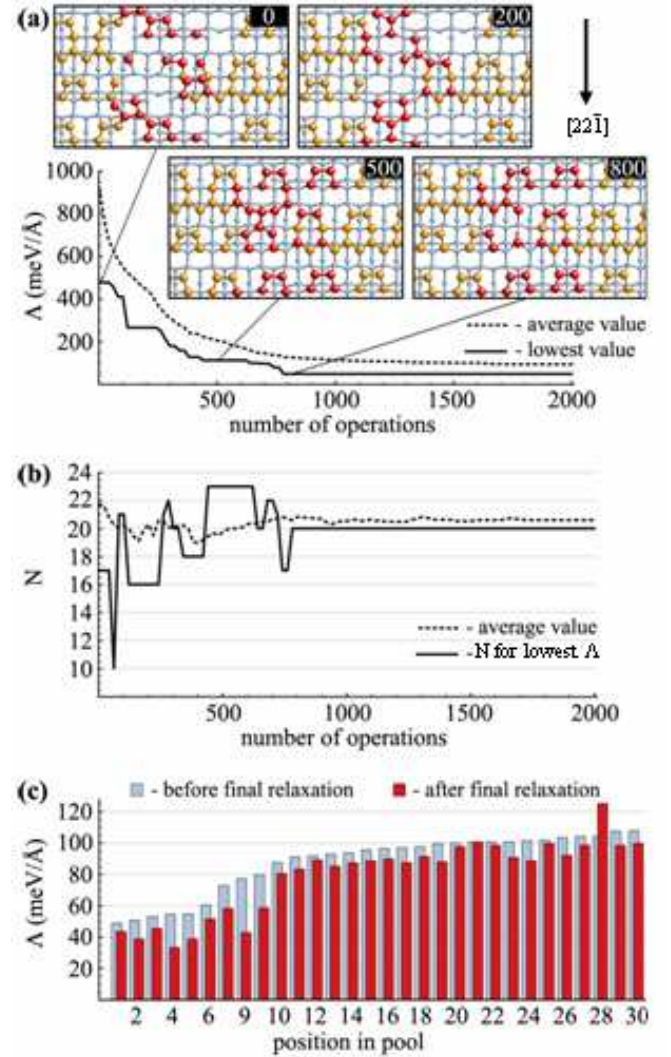}
\caption{(Color online) Finding the structure of $[22{\overline
1}]$-normal steps \cite{geosimcell} on Si(114). (a) Step energy
$\Lambda$ of the lowest-energy structure (solid line) and averaged
across the pool (dashed line) during the genetic evolution. The
lowest energy structure is shown after 0, 200, 500, and 800
crossover operations; the atoms subjected to optimization are
shown as darker spheres in the insets. (b) Evolution of the
average number of atoms across the pool (dashed line) and of the
atom number corresponding to lowest-energy member (solid line).
(c) Step energies before and after the final relaxation of all
members of the genetic pool. The formation energies decrease upon
full relaxation unless bonds are broken in the process (as found
in the case of structure no. 28). } \label{lambda-N-lambdafinal}
\end{figure}

\begin{figure*}
   \includegraphics[width=15cm]{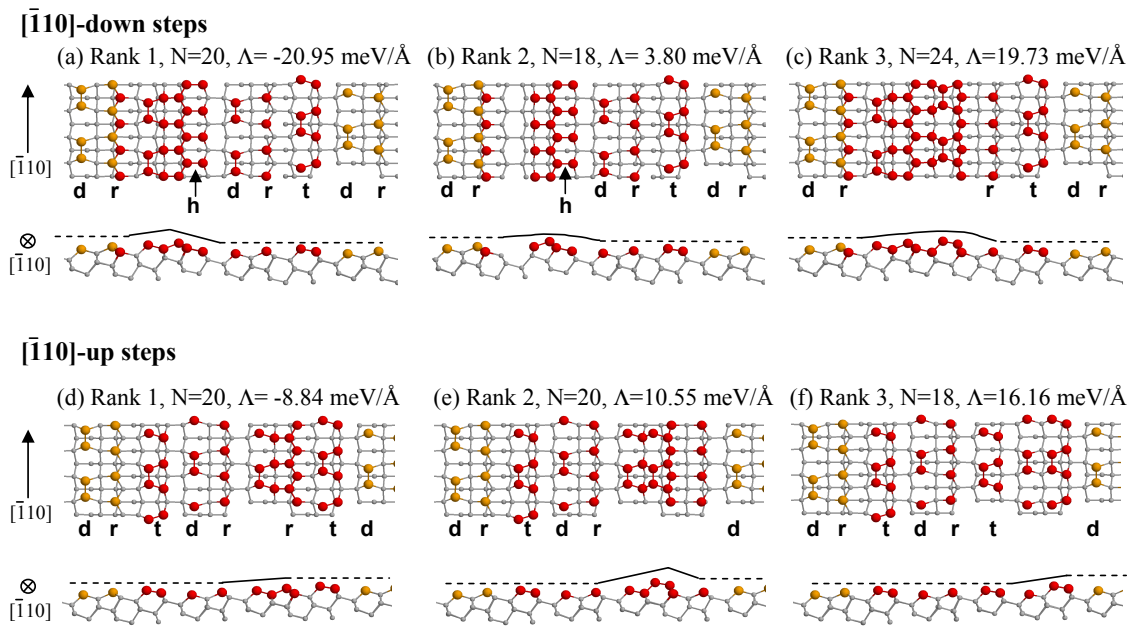}
 \caption{(Color online) Low-energy step structures of
[${\overline 1}10$]-oriented
 steps on the Si(114) surface. The atoms subjected to optimization
 are represented by dark spheres, while the atoms making up the
 terrace reconstructions are the lighter ones. The remaining atoms
 are shown as smaller gray spheres. The structural motifs on the terraces are rows of
 dimers (d), rebonded atoms (r), and tetramers (t). Some or all of these
 motifs also make up the shown step structures with the exception of the most favorable
 down-step models [panels (a) and (b)] which include hexagon rows denoted by ``h''.
 A schematic contour of the step topology was included in each side view to aid the eye.}  \label{top3-110steps}
\end{figure*}

The methodology that we choose for finding the step structures is
based on a genetic algorithm, which has been shown to achieve fast
convergence using aggressive multi-particle moves for systems of
any dimensionality from clusters to bulk crystalline materials
(see, e.g., Refs. \onlinecite{prl-clusters, nature-clusters,
prl-nl-nanowires, ga-surface, hjwz-perspective, ss-si114-global,
ga3d}). In a rather simple but efficient way, the algorithm
simulates a biological process in which a set of individuals
evolves with the goal of producing fit children, i.e. new step
structures with low formation energies. For all runs we have
systematically kept a pool of $p=30$ atomic structures and
subsequently tested that a range of $30 \leq p < 100$ is
appropriate for this problem, i.e. the pool is large enough that
the evolution is not likely to get stuck in a local minimum and is
sufficiently small so that the cross-over moves are not wasted
optimizing mostly high-energy local minima. In general, the size
of the pool should be determined by numerical experimentation for
the particular system under study. The starting $p$ structures
(``Generation Zero") are simply collections of atoms that are
randomly positioned in the step region (colored gray in
Fig.~\ref{boundary-mating}a) then relaxed to nearest local minima
of the potential energy of the system. The evolution from the
current generation to the next one occurs via crossover processes
in which the step structures corresponding to two randomly chosen
parents $A$ and $B$ from the pool are combined to create a new
(child) structure, $C$. Referring to
{Fig.~\ref{boundary-mating}b}, the crossover of parents $A$ and
$B$ is achieved by sectioning them with the same random plane,
then retaining atoms from each parent located on different sides
of this plane to create the child $C$. The plane is chosen here to
be parallel to [114] (any polar angle about this direction is
allowed) and located so as to pass through a randomly chosen point
in the rectangle projected by the step zone onto the (114) plane.
The operation so defined ({Fig.~\ref{boundary-mating}b}) has
built-in potential to generate child structures with different
numbers of atoms than their parents. Any child that is {\em
structurally distinct from all pool members} is considered for
inclusion in the genetic pool based on its formation energy per
unit length, which should be lower than that of the highest ranked
(i.e. least favorable) member of the pool. To preserve the total
population, the structure with the highest formation energy is
discarded upon inclusion of a new child. In a genetic algorithm
run, the crossover operation is repeated to ensure that the lowest
energy structure of the pool has stabilized; as such, the present
systems require 2000 operations as illustrated by the curves in
Fig.~\ref{lambda-N-lambdafinal}a which become flat as the
algorithm progresses.

\begin{figure}
   \includegraphics[width=8.5cm]{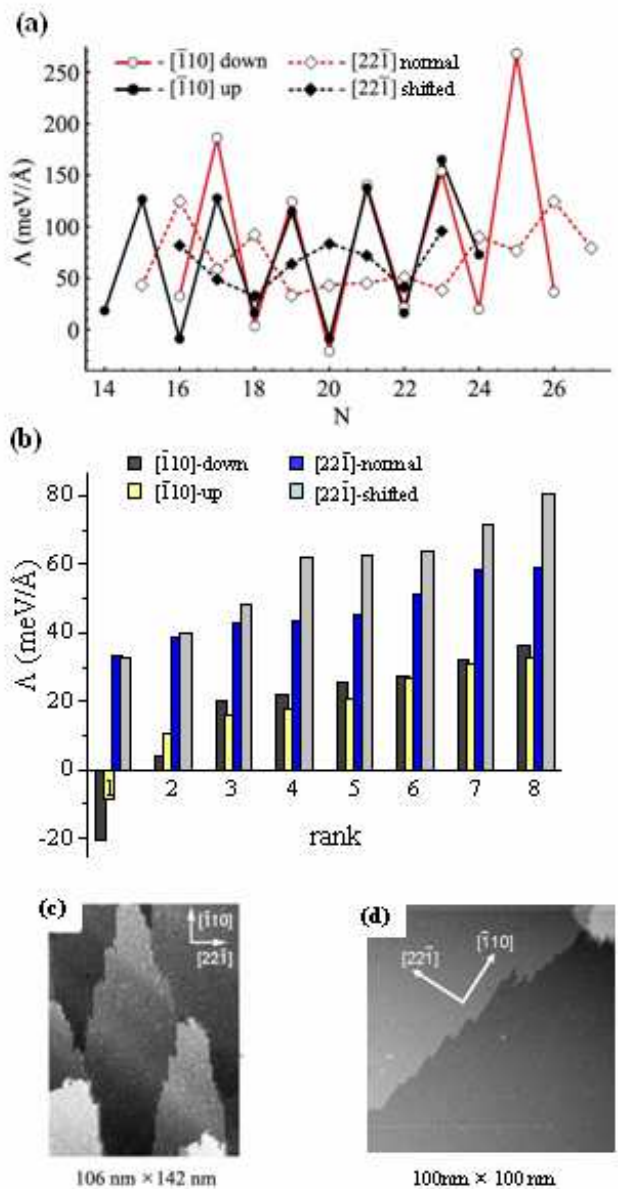}
 \caption{(Color online) Results of the genetic algorithm (a,b) and experimental
 observations of steps on Si(114) surfaces (c,d).
(a) Lowest step energies attained at various atom numbers in the
step zone.  %
(b) Final step formation energies of the top-ranking structures
for each of the four types of steps described in text. The
[${\overline 1}$10] steps have consistently lower energies
than similarly ranked [22${\overline 1}$] structures. %
(c) STM image of a vicinal Si(114) surface obtained after cleaning
and brief annealing (reproduced from Ref.\onlinecite{Si114-apl}
with permission of the American Institute of Physics). (d) STM
image taken after flashing at 1225$^{\text o}$C followed by 30
minute annealing at 450$^{\text o}$C (courtesy of D. E. Barlow, A.
Laracuente, and L.J. Whitman). These experimental observations
show that the $[{\overline 1}10]$ steps are preferentially longer
than the [22${\overline 1}$] steps.} \label{magic-energies-exp}
\end{figure}

The formation energy of a step structure is defined as a
per-length quantity that is in excess to the bulk and surface
energies,\cite{shifted-pbc} and therefore can be written as
\begin{equation}
 \Lambda = \frac{1}{L_y}(E_m - N_m e_b - \gamma  A), \label{eq:formation}
\end{equation}
where $E_m$ is the total energy of the $N_m$ atoms that are
allowed to move within a projected area $A=L_x L_y$ with the
dimension $L_x$ ($L_y$) perpendicular (parallel) to the step,
$e_b$ is the bulk cohesion energy of Si, and $\gamma$ is the
surface energy of the flat Si(114) surface. The potential we have
chosen to model the atomic interactions is the one developed by
Lenosky et al.,\cite{hoep} because it has shown reasonable
transferability for diverse atomic environments present on
high-index Si surfaces.\cite{{ga-surface},ss-si114-global} If all
the atoms of the supercell are allowed to move when calculating
the formation energy [Eq.~(\ref{eq:formation})], then each update
of the genetic pool will be too slow for the algorithm to be
practical. On the other hand, if we only relax the atoms in the
step zone, then Eq.~(\ref{eq:formation}) will include not only the
formation energy but also the elastic interactions of the step
with the nearby rigid boundaries of the step region. To reach a
good compromise between the full accuracy of
Eq.~(\ref{eq:formation}) (which would be achieved when {\em all}
atoms in the supercell are relaxed) and the speed required to sort
out many structures per unit time, we introduce a padding zone
that is relaxed along with the step region while the keeping the
reconstructed support zone fixed (refer to
Fig.~\ref{boundary-mating}b). The use of the padding zone will
diminish the elastics interactions in Eq.~(\ref{eq:formation})
while keeping the number of atoms that are relaxed after each
crossover small.  At the end of any genetic algorithm run, a full
relaxation (all atoms allowed to move) is performed for all pool
members in order to refine the step energies.

Typical results of the genetic algorithm for steps are shown in
{Fig.~\ref{lambda-N-lambdafinal}a}, which displays the evolution
of the lowest and of the average formation energy of the genetic
pool as a function of the number of crossover operations. Both the
lowest and the average formation energies show a rapid decrease at
the beginning of the evolution, followed by slower decay in the
later stages. The lowest-energy $[22{\overline 1}]$-normal
configuration was found in less than 1000 operations, and has been
retrieved in four runs started from different Generation Zero
scenarios with no significant change in the total number of
crossover moves. Since the crossover operation described above
creates new structures with variable numbers of atoms, the number
of atoms $N$ in the step region is optimized at the same time as
the atomic positions ($N$ is always smaller than the total number
of atoms allowed to relax, $N_m$). Illustrative of the search for
the optimal particle number $N$ is
{Fig.~\ref{lambda-N-lambdafinal}b}, which displays the evolution
of the average number of atoms in the genetic pool and the
particle number corresponding to the lowest-energy member.
{Fig.~\ref{lambda-N-lambdafinal}c} shows that upon final full
relaxation a certain amount of energetic reordering does occur,
but this reordering is merely a refinement and does not eliminate
from consideration any of the best structures deemed favorable
prior to the complete relaxation of all step models in the pool.
When using the algorithm for an arbitrary line defect, the
formation energy comparison before and after final relaxation
offers the most useful criterion for adjusting the size of the
padding zone so as to provide sufficient relaxation without
rendering the calculations intractable.

The best three structures for the up- and down-steps oriented
along $[{\overline 1}10]$ are shown in {Fig.~\ref{top3-110steps}},
along with their formation energies after the final relaxation and
their optimal atom numbers $N$. The most favorable up-step and
down-step both have negative formation energies, which is a known
artifact of the empirical potentials.\cite{shifted-pbc,
si001review} Without placing undue significance on the negative
sign, we focus on the ranking of the formation energies and the
corresponding structures. The reconstruction of the flat Si(114)
surface consists of rows of dimers (d), rebonded atoms (r) and
tetramers (t) in this specific periodic sequence
(...-d-r-t-d-r-t-d-....) along the $[22{\overline 1}]$ direction.
\cite{Si114-prl} Since we allowed for a large width of the step
region,\cite{largewidth} the steps can negotiate their width and
location during the genetic evolution. This is apparent in
{Fig.~\ref{top3-110steps}}, which shows that the sequence of
motifs (d, r, t) is continued seamlessly from each terrace into
the step zone until the atomic structure and the location of the
step edge are determined. The best $[{\overline 1}10]$ down-step
structures (Fig.~\ref{top3-110steps}a,b) include a row of hexagons
(labelled by ``h" in Fig.~\ref{top3-110steps}) in addition to
motifs already encountered on the terraces. Other low-energy steps
are observed to simply consist of a gap in the -d-r-t- sequence of
the structural features on the terraces. For example,
Fig.~\ref{top3-110steps}c shows a down step that contains dimers,
rebonded atoms and tetramers in the correct order, but which are
bonded to the upper (lower) terrace via elimination of one
tetramer (dimer) row from the -d-r-t- sequence on the terraces.
The most favorable up-step structures contain only rows of dimers
and rebonded atoms (Fig.~\ref{top3-110steps}d), all motifs in a
different order (d-t-r, Fig.~\ref{top3-110steps}e), or only rows
of dimers and tetramers (Figs.~\ref{top3-110steps}f). The
algorithm is thus able to find narrow steps
({Fig.~\ref{top3-110steps}d}), wide ones
({Fig.~\ref{top3-110steps}c}), as well as steps with intermediate
widths ({Fig.~\ref{top3-110steps}a,b,e,f}). This morphological and
structural diversity is a tell-tale sign of the superior
configuration sampling achieved here with just one simple genetic
operation (the crossover, Fig.~\ref{boundary-mating}b).
\cite{mutations}

To provide a closer look at the way the algorithm sorts through
different numbers of atoms, we have plotted  the lowest formation
energy found for every number of atoms $N$ attained {\em during}
the evolution ({Fig.~\ref{magic-energies-exp}a}). Such a plot
shows that the algorithm can visit, in the same evolution, several
structures of particularly low formation energies (magic-number
atomic configurations) and select them as part of the genetic
pool. Magic structures are found for even values of $N$ both for
the up and down $[{\overline 1}10]$ steps, as seen in
{Fig.~\ref{magic-energies-exp}a}. The same figure shows that the
formation energy of $[22{\overline 1}]$-normal steps also has a
few distinct local minima, which are located at odd values of $N$.
On the other hand, the $[22{\overline 1}]$-shifted configuration
does not have magic number behavior. In this case, there are two
deep minima for the range of atom numbers spanned, but after the
final relaxation they are found to have the same structure only
translated by one full terrace unit cell \cite{Si114-prl} along
the $[{\overline 1}10]$ axis.

Finally, in {Fig.~\ref{magic-energies-exp}b} we report the
formation energies of the lowest eight structures in the pool for
each of the four configurations studied.  The figure shows that
the steps oriented in the $[{\overline 1}10]$ direction have
smaller formations energies than those along $[22{\overline 1}]$
for all top ranking structures found. For the $[{\overline 1}10]$
direction the down-steps are easier to form than the up-steps,
while for the $[22{\overline 1}]$ direction the up and down steps
have identical structures and energies. The conclusion that
$[{\overline 1}10]$ steps are more favorable than $[22{\overline
1}]$ steps is consistent with the general expectation that a
direction of higher symmetry (i.e. $[{\overline 1}10]$) yields
lower step energies than a low symmetry one. Ideally, these
results could be tested by recalculating the formation energy of
steps at the level of density functional theory (DFT)
calculations. While in principle DFT level calculations of
formation energies of steps on Si(114) are tractable, the number
of atoms involved is at least several hundred atoms which is
beyond our present computational means.

However, we have found that existing experimental observations do
support our genetic algorithm results, albeit qualitatively.
Laracuente et al.\cite{Si114-apl} reported STM images (reproduced
here in Fig.~\ref{magic-energies-exp}c) in which the [${\overline
1}$10] steps are clearly preferred over the [22${\overline 1}$]
ones even when, due to the preparation conditions\cite{Si114-apl}
steps may not assume the very lowest-energy structure for given
direction and terrace configuration: this is consistent with the
simulation results in Fig.~\ref{magic-energies-exp}b which show
that [${\overline 1}$10] steps are lower in energy for {\em
metastable} structures ranked within the first eight at the end of
the genetic evolution. More recently, Whitman and coworkers have
also imaged step configurations after long anneals at 450$^{\text
o}$C. These recent measurements (shown in
Fig.~\ref{magic-energies-exp}d) are more likely to correspond to
lowest-energy step structures, and again show that the
$[{\overline 1}10]$ are longer (more favorable) thus lending
support to our simulation results. In terms of atomic positions,
so far there has been no proposal for the structure of steps on
Si(114). We have proposed here several low-energy structures
({Fig.~\ref{top3-110steps}}) found via the genetic procedure,
structures which are amenable to experimental testing via
high-resolution STM measurements combined with ab initio density
functional calculations.

The versatile variable-number algorithm described in previous
reports on two-dimensional surface systems \cite{ga-surface,
ss-si114-global} has recently been extended to the case of 3-D
crystal structure prediction.\cite{ga3d} While in the 2-D and 3-D
cases structure prediction methods were already
available\cite{alt-2d, alt-3d} prior to the introduction of the
genetic approaches,\cite{ga-surface,ga3d} to our knowledge the
present variable-number genetic algorithm is the first robust
approach to atomic structure determination for surface-supported
1-D systems. A reader may rightfully argue that the use of
empirical potentials could cast doubts on the results obtained by
using this algorithm. Although empirical (or even tight-binding)
potentials do have artifacts which lead to spurious minima, these
minima may  be accommodated to some extent by increasing the size
of the pool. What makes the algorithm robust is not the specific
potential model, but rather the concept of an evolved database
whose optimized members (with different atom numbers and different
structures) can be studied subsequently at any level of theory,
including ab initio calculations. It is worth noting that, in
fact, the reliance on empirical potential is much smaller in the
present genetic algorithm than in the case of molecular dynamics
and continuous-space Monte Carlo, \cite{alt-2d, alt-3d} because in
the current implementation the genetic algorithm only relies on an
acceptable accuracy of the local minima energies without the
additional requirement of a good description of the height of the
barriers between these minima.

In conclusion, we have presented a general way to determine the
structure of steps on reconstructed crystal surfaces, and applied
it to find the structure of monatomic steps on Si(114)-$2\times
1$. A key finding of this paper is that the step structure problem
can be solved by what is probably the simplest genetic algorithm,
i.e. an algorithm which is based on greedy selection of new
members structurally distinct from the old ones and on a single
real-space operation (cross-over between two parents using planar
cuts). We speculate that the reason for which this simple approach
works is that the underlying bulk crystal provides a strong
template onto which the structure can relax while also obeying the
periodic boundary conditions imposed. The bulk template lowers the
number of both the distinct and the symmetry-equivalent
configurations that the pool members can visit, and thus decreases
the complexity of the problem as compared, e.g., with the
optimization of an atomic cluster with the same range for the
number of atoms. The variable-number approach is particularly
well-suited for other problems beyond that of single-height steps
on clean high-index Si surfaces. Structure of the edge between two
stable facets that bound a quantum dot, gold atom decorations of
stepped surfaces, \cite{553-Au-Spain,553-Au-Wisc,metaldecoration}
adsorbate-induced surface reconstructions,\cite{ad-inducedreco}
steps on compound semiconductor surfaces,\cite{gaas-steps} and
structure of step bunches during growth (e.g.
Refs.~\onlinecite{105-bunching-SS,edw}) -- to give a few
significant examples -- can be studied systematically using the
procedure presented here, with the only necessary modifications
concerning the geometry of the supercell and the expression of the
formation energy to account for a second atomic species.

{\em Acknowledgment.} CVC thanks Lloyd Whitman of the Naval
Research Laboratory for very useful discussions and for graciously
providing recent STM images of steps on Si(114) surfaces. We
gratefully acknowledge the support of the National Center for
Supercomputing Applications at Urbana-Champaign through grant no.
DMR-050031.


\begin{thebibliography}{99}
\bibitem{largewidth} The size of the step zone was chosen so as
to cover one full reconstructed unit cell on each side of the
(initially arbitrary) step location.
\bibitem{mutations}In most genetic algorithm implementations, mutations (small random displacements
of arbitrarily selected atoms) are necessary to provide paths for
the pool members to evolve towards global minima or towards
stationary states \cite{gabook}. Due to the requirement to allow
only structurally distinct pool members at all times, mutations
turned out to be unnecessary in the present algorithm.
\end{thebibliography}
\end{document}